\documentclass[apl,a4paper,twocolumn,floatfix,showpacs,showkeys,amsmath,amssymb,nobibnotes,altaffilletter]{revtex4}

\usepackage{graphicx}
\usepackage{pslatex}
\usepackage{xspace}
\usepackage{subfigure}
\usepackage[latin1]{inputenc}
\usepackage[T1]{fontenc}
\usepackage{amsmath}

\newcommand{\pht}{poly(3-hexyl thiophene)\xspace}
\newcommand{\pcbm}{[6,6]-phenyl-C$_{61}$ butyric acid methyl ester\xspace}
\newcommand{\pedot}{poly(3,4-ethylenedioxythiophene):poly(styrenesulfonate)\xspace}
\newcommand{\ppv}{poly[2-methoxy-5-(3$'$,7$'$-dimethyloctyloxy)-1,4-phenylenevinylene]\xspace}
\newcommand{\degree}{$^{\circ}$}

\begin{document}

\title{Direct and charge transfer state mediated photogeneration in polymer--fullerene bulk heterojunction solar cells}

\author{M.~Mingebach$^{1}$}
\author{S.~Walter$^{1}$}
\author{V.~Dyakonov$^{1,2}$}
\author{C.~Deibel$^{1}$}\email{deibel@physik.uni-wuerzburg.de}
\affiliation{$^{1}$ Experimental Physics VI, Julius-Maximilians-University of W{\"u}rzburg, 97074 W{\"u}rzburg, Germany}
\affiliation{$^{2}$ Bavarian Centre for Applied Energy Research (ZAE Bayern), 97074 W{\"u}rzburg, Germany}

\date{\today}

\begin{abstract}
We investigated photogeneration yield and recombination dynamics in blends of \pht (P3HT) and \ppv (MDMO-PPV) with \pcbm (PC$_{61}$BM) by means of temperature dependent time delayed collection field (TDCF) measurements. 
In MDMO-PPV:PC$_{61}$BM we find a strongly field dependent polaron pair dissociation which can be attributed to geminate recombination in the device. Our findings are in good agreement with field dependent photoluminescence measurements published before, supporting a scenario of polaron pair dissociation via an intermediate charge transfer (CT) state. 
In contrast, polaron pair dissociation in P3HT:PC$_{61}$BM shows only a very weak field dependence, indicating an almost field independent polaron pair dissociation or a direct photogeneration.
Furthermore, we found Langevin recombination for MDMO-PPV:PC$_{61}$BM and strongly reduced Langevin recombination for P3HT:PC$_{61}$BM.

\end{abstract}



\maketitle

%-------------------Introduction ------------------------------------------------------------
The high interest in organic photovoltaics lead to rapidly rising solar cell efficiencies during the last years, reaching 8.3\%~\cite{Green2011} for a single junction device and even above 10\%~\cite{Green2012} for tandem architectures. 
Despite these impressive achievements, improvements of device efficiency and lifetime remain still high on the agenda. An important step for further optimizations is the understanding of the device performance limiting processes, such as charge photogeneration and recombination~\cite{Deibel2010, Brabec2010}. A fast generation of free charge carriers with high yield is desirable to ensure high power conversion efficiencies. The dissociation of singlet excitons into free charge carriers may occur either directly~\cite{Howard2010} or via an intermediate step involving Coulomb bound charge transfer (CT) states~\cite{Deibel2010c, Bakulin2012}. The latter pathway may have serious drawbacks, e.g., due to recombination of charge pair states to long-lived lower lying electronic states~\cite{Nuzzo2010, Liedtke2011}. The presence or absence of intermediate states becomes evident in the electric field dependence of free charge carrier generation. The photogeneration and recombination dynamics can be investigated by time-delayed collection field (TDCF) measurements. This method was introduced by Mort~\cite{Mort1980} in 1980 who investigated geminate and nongeminate recombination in amorphous silicon. During the last years, it has also been applied to organic bulk heterojunction (BHJ) solar cells to study the generation, recombination and lifetime of charge carriers~\cite{Offermans2006, Kniepert2011}.

Here, we investigated organic BHJ solar cells containing blends of either MDMO-PPV:PC$_{61}$BM or P3HT:PC$_{61}$BM by temperature dependent TDCF measurements. We observed a strongly field dependent polaron pair dissociation in blends of MDMO-PPV:PC$_{61}$BM, which is in  good agreement with its CT exciton binding energy of up to 200 meV, determined by photoluminescence measurements~\cite{Kern2011,Hallermann2008}. In contrast, an almost field and temperature independent charge collection was experimentally found in P3HT:PC$_{61}$BM. 
In addition, we discuss the nongeminate recombination dynamics in both blends and distinguish between Langevin recombination in MDMO-PPV:PC$_{61}$BM and a strongly reduced Langevin recombination in P3HT:PC$_{61}$BM.

%----------------------------Experimental------------------------------------------------------------------
All samples presented here were processed on indium tin oxide (ITO) covered glass substrates on which we spincast a thin layer of \pedot (PEDOT:PSS). In a subsequent step the active layer consisting of either P3HT:PC$_{61}$BM or MDMO-PPV:PC$_{61}$BM (1:1 and accordingly 1:4 weight ratio dissolved in chlorobenzene) was applied. The active layer thicknesses were 155 nm and 90 nm, respectively, as determined by a Veeco Dektak 150 profilometer. Before the the final step of evaporating metal contacts consisting of Ca (3 nm) and Al (100 nm), the devices containing P3HT:PC$_{61}$BM were thermally annealed on a hotplate for 10 minutes at 130 \degree{C}. The studied samples had active areas of 3 mm$^{2}$ and exhibited power conversion efficiencies of 3.2\% (P3HT:PC$_{61}$BM) and 1.6\% (MDMO-PPV:PC$_{61}$BM) with fill factors of 66\% and 53\%, respectively. To determine the photovoltaic performance, we used a Xe arc lamp which was adjusted to standard testing conditions~\cite{Shrotriya2006}. All steps from applying the active layer to the photovoltaic characterization were performed in a glovebox system under N2 atmosphere. The devices were transferred to a He cryostat for temperature dependent TDCF measurements.

For charge carrier generation in TDCF measurements, we used short laser pulses (< 80 ps) of a 532 nm neodymium-doped yttrium aluminum garnet (Nd:YAG) laser with a low repetition rate of 5 Hz to prevent electrical charging of the device. The device under test was either kept at a constant prebias (-5 V < $V_{pre}$ < 1 V) for a variable delay time (150 ns < $t_{delay}$ < 100 $\mu s$) or, vice versa, the delay time was kept constant while varying the prebias. Right after the delay, all remaining charges were extracted by a negative collection voltage ($V_{col}$ = -6 V) to avoid recombination losses during charge extraction. To account for capacitive effects, we corrected all measurements by subtracting the corresponding extraction current transient taken without laser pulse. 

The photocurrent response measured is a displacement current displaying two peaks: The first one is due to the prebias $V_{pre}$ during the delay time and the second one due to the collection voltage $V_{col}$ during charge extraction. Hence, the integral over these two peaks, one corresponding to the charge collected during the prebias $Q_{pre}$ and the other one to the charge extracted during collection bias $Q_{col}$, is expected to be equal to the sum of all charge displacements $Q_{tot}$. 

For low laser pulse fluences we observed a linearly increasing $Q_{col}$ with increasing illumination intensity ($V_{pre}$ = 0 V). As previously shown by Kniepert et al.~\cite{Kniepert2011} such behavior indicates the absence of nongeminate recombination in the active layer of the device. We therefore selected this regime for our studies of field dependend photogeneration. At higher illumination intensities, the extracted charge density increases only sublinearly, as photogenerated charge carriers are lost due to nongeminate recombination and can no longer be extracted by the collection field.

%--------------Results and Discussion-------------------------------

\begin{figure}
	\includegraphics[width=8cm]{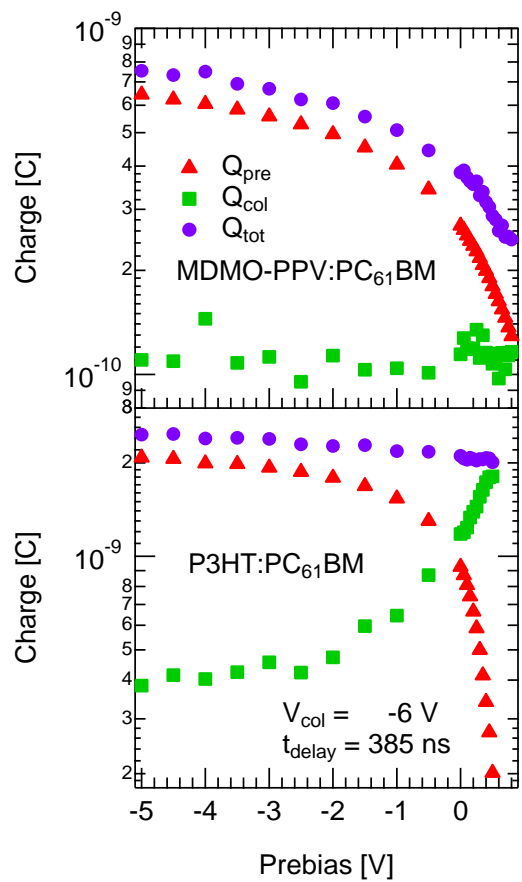}
	\caption{(Color online) Integrated charge from field dependent TDCF measurements. The MDMO-PPV:PC$_{61}$BM 1:4 blend (top) shows a strong dependence on the electric field due to geminate recombination, indicating charge separation via a CT state. In contrast, the P3HT:PC$_{61}$BM 1:1 (bottom) blend shows a low field dependence and therefore almost no influence of geminate recombination on charge separation.}
	\label{fig:prebias}
\end{figure}

%--------------------Geminate Recombination ------------------------------------------------------------------
Taking these findings into account, we performed field and temperature dependent TDCF measurements in blends of MDMO-PPV:PC$_{61}$BM and P3HT:PC$_{61}$BM. The delay time (385 ns) and collection voltage (-6 V) were set constant and the prebias was varied between -5 V and the individual open circuit voltage of the respective device. The corresponding results are shown in Figure~\ref{fig:prebias}. 

In MDMO-PPV:PC$_{61}$BM (Fig. \ref{fig:prebias}, top) we observed a strong field dependence of $Q_{tot}$. The saturation value at $V_{pre}$ = -5 V decreases with applied field reaching 70\% of the initial value for zero internal field under open circuit conditions. At high negative fields exceeding the Coulomb attraction potential, the influence of geminate recombination is negligible. As we already ruled out the influence of nongeminate recombination for negative prebias by measuring at very low light intensities, the field dependence observed in Fig.~\ref{fig:prebias} (top) is mainly due to the field dependent charge photogeneration. The photocurrent becomes more strongly field dependent at low internal fields, i.e.\ under positive prebias close to or under open circuit conditions, which is due to a combination of field dependent charge photogeneration in addition to nongeminate recombination. In general, the field dependent photogeneration is in good agreement with charge separation via charge transfer (CT) states, accounting for a considerable CT exciton binding energy of up to 200 meV, as determined by photoluminescence measurements~\cite{Hallermann2008, Kern2011}.

In contrast, P3HT:PC$_{61}$BM (Fig. \ref{fig:prebias}, bottom) shows a higher amount of charge and only a rather low field dependence of about 20\%  for $Q_{tot}$ between 0 and -5V, and probably less than 10\% between open and short circuit conditions. Previously, a field dependence of approx.\ 15\% between open and short circuit conditions was reported based on Monte Carlo simulations~\cite{Deibel2009a} and photocurrent experiments~\cite{Limpinsel2010}. Recently, transient absorption measurements indicated fast direct generation of free charge carriers with a fraction of about 15\% geminate recombination in a P3HT:PCBM thin film under open circuit conditions~\cite{Howard2010}. This high yields of photogeneration could point at an efficient polaron pair dissociation route via excited, or hot, CT complexes~\cite{Deibel2010c}. Bakulin et al.~\cite{Bakulin2012} proposed that the field dependence of photogeneration is given by the energy needed to reach delocalized band states.

\begin{figure}
	\includegraphics[width=8cm]{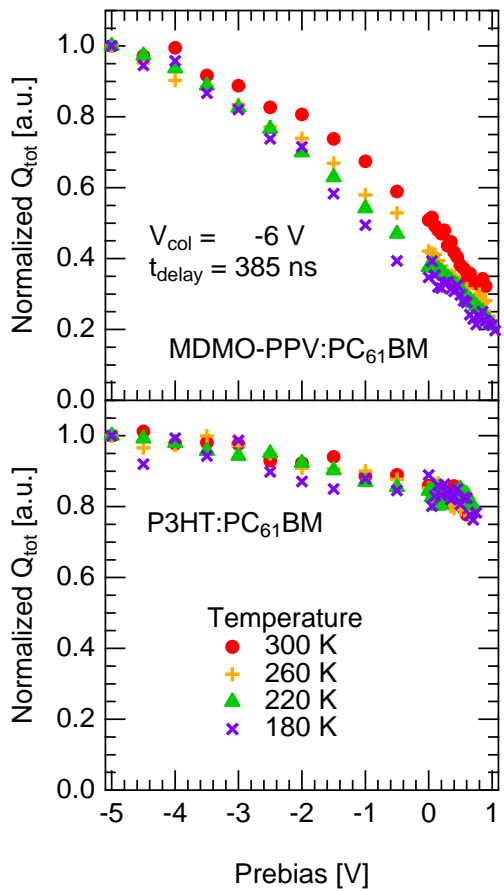}
	\caption{(Color online) Integrated charge from field and temperature dependent TDCF measurements. In the case of MDMO-PPV:PC$_{61}$BM (top) a clear temperature dependence of the field dependent polaron pair dissociation is observable, indicating a thermally activated process. In contrast, no temperature dependence of the field dependent charge separation can be found for P3HT:PC$_{61}$BM (bottom).}
	\label{fig:Tdep}
\end{figure}

Temperature dependent measurements for both material systems are presented in Fig.~\ref{fig:Tdep}. For sake of comparability, the data was normalized to $Q_{tot}$(-5V), although not all curves at low temperatures were saturated. Nevertheless, the temperature dependent effect for MDMO-PPV:PC$_{61}$BM (Fig. \ref{fig:Tdep}, top) is clearly observable. The field dependence of the photogeneration increases for decreasing temperatures and therefore accounts for a thermally activated process. This further supports the scenario of polaron pair dissociation via a CT state, as the binding energy of the CT exciton is more difficult to overcome at low temperatures. In contrast, in P3HT:PC$_{61}$BM (Fig. \ref{fig:Tdep}, bottom) no temperature dependence of the already weak field dependence can be found, consistent with a barrier-free photogeneration as noted above.

\begin{figure}
	\includegraphics[width=8cm]{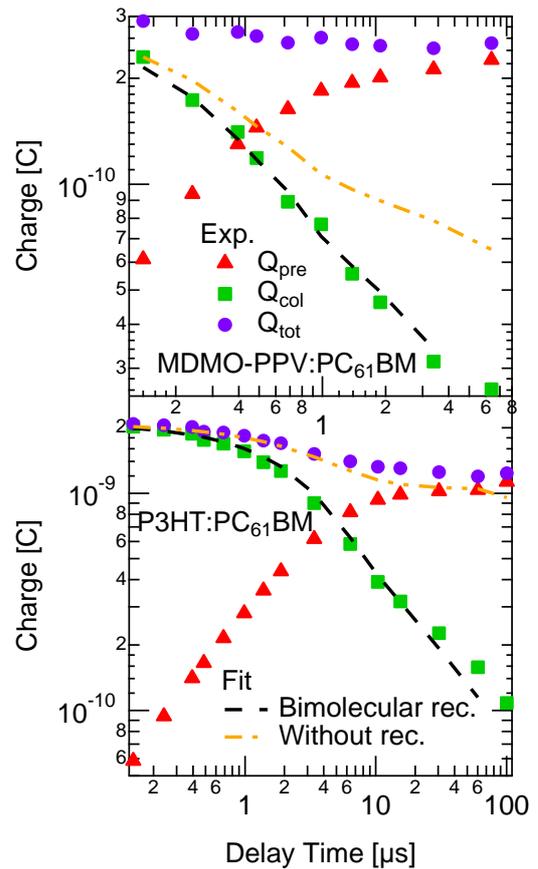}
	\caption{(Color online) Integrated charge from TDCF measurements of MDMO-PPV:PC$_{61}$BM (top) and P3HT:PC$_{61}$BM (bottom) at room temperature for different delay times. Q$_{tot}$ decreases for increasing delay times due to bimolecular recombination. The iterative fit forward in time to Q$_{col}$ is solely based on bimolecular recombination (black dashed line) and results in the Langevin recombination prefactor.}
	\label{fig:p200delay}
\end{figure}

%----------------------Nongeminate recombination ------------------------------------------------------------------
To study nongeminate recombination in both blends, we performed TDCF measurements with a variable delay time of 150 ns up to 100 $\mu s$. Meanwhile prebias and collection field were set constant. The corresponding room temperature data for MDMO-PPV:PC$_{61}$BM and P3HT:PC$_{61}$BM is shown in Fig.~\ref{fig:p200delay}. The total amount of charges $Q_{tot}$ decreases with increasing delay time due to bimolecular recombination of the photogenerated charges, as expected. To extract the Langevin recombination prefactor $\gamma$ we applied an iterative fit forward in time (dashed line) to $Q_{col}$ similar to the procedure applied by Kniepert et al.~\cite{Kniepert2011}. The model referred to is solely based on bimolecular recombination and in good agreement with the experimental data in the time range studied. The yellow (color online) dash-dotted line was calculated by the same model while accounting only for extraction, neglecting the bimolecular recombination. The Langevin recombination prefactor is defined as~\cite{Pivrikas2005, Deibel2009}
\begin{equation}
\gamma = \zeta \frac{q}{\varepsilon}\mu
\end{equation}
where $q$ is the elementary charge, $\varepsilon$ the dielectric constant (here approximated with 3.5) and $\mu$ the average electron and hole mobility. 
Concerning MDMO-PPV:PC$_{61}$BM, we extracted a Langevin recombination prefactor of $\gamma$~=~8.2~$\cdot$~10$^{-17}~\frac{m^{3}}{s}$. With $\mu$~=~3~$\cdot$~10$^{-8}~\frac{m^{2}}{Vs}$~\cite{Dennler2006}, a reduction factor of $\zeta \approx$~0.5 is found, which is close to unity. This indicates the Langevin recombination at room temperature in blends of MDMO-PPV:PC$_{61}$BM and is consistent with literature~\cite{Mozer2005}. 

In P3HT:PC$_{61}$BM, we determined  $\gamma$~=~5.3~$\cdot$~10$^{-18}~\frac{m^{3}}{s}$, which is in good agreement with literature~\cite{Kniepert2011}. With $\mu$~=~5.5~$\cdot$~10$^{-8}~\frac{m^{2}}{Vs}$~\cite{Koster2006b, Baumann2008} we found a very low reduction factor of $\zeta \approx$0.02. In contrast to MDMO-PPV:PC$_{61}$BM we observe a strongly reduced Langevin recombination in P3HT:PC$_{61}$BM, again in good agreement with previous works~\cite{Pivrikas2005, Deibel2009}.

%--------------------Conclusions ------------------------------------------------------------------
To conclude, we studied charge carrier generation and collection yield in blends of MDMO-PPV:PC$_{61}$BM and P3HT:PC$_{61}$BM by temperature dependent TDCF experiment. In MDMO-PPV:PC$_{61}$BM, we found a strong field dependence of the polaron pair dissociation, which becomes even more pronounced at lower temperatures, indicating polaron pair dissociation via CT states. In contrast, only a weak field and no temperature dependence was found for P3HT:PC$_{61}$BM. These findings can be explained within the scenario of direct generation of free charge carriers or polaron pair dissociation via (hot) CT states in the respective blend system. Furthermore, we determined the nongeminate recombination prefactors and conclude on a Langevin recombination in MDMO-PPV:PC$_{61}$BM and a strongly reduced Langevin recombination in the case of P3HT:PC$_{61}$BM.  

%-------------------Acknowledgments ------------------------------------------------------------------

The current work is supported by the Bundesministerium f{\"u}r Bildung und Forschung in the framework of the Grekos Project (Contract No. 03SF0356B). C.D. gratefully acknowledges the support of the Bavarian Academy of Sciences and Humanities. V.D.'s work at the ZAE Bayern is financed by the Bavarian Ministry of Economic Affairs, Infrastructure, Transport, and Technology.


\begin{thebibliography}{23}
\expandafter\ifx\csname natexlab\endcsname\relax\def\natexlab#1{#1}\fi
\expandafter\ifx\csname bibnamefont\endcsname\relax
  \def\bibnamefont#1{#1}\fi
\expandafter\ifx\csname bibfnamefont\endcsname\relax
  \def\bibfnamefont#1{#1}\fi
\expandafter\ifx\csname citenamefont\endcsname\relax
  \def\citenamefont#1{#1}\fi
\expandafter\ifx\csname url\endcsname\relax
  \def\url#1{\texttt{#1}}\fi
\expandafter\ifx\csname urlprefix\endcsname\relax\def\urlprefix{URL }\fi
\providecommand{\bibinfo}[2]{#2}
\providecommand{\eprint}[2][]{\url{#2}}

\bibitem[{\citenamefont{Green et~al.}(2011)\citenamefont{Green, Emery,
  Hishikawa, and Warta}}]{Green2011}
\bibinfo{author}{\bibfnamefont{M.~A.} \bibnamefont{Green}},
  \bibinfo{author}{\bibfnamefont{K.}~\bibnamefont{Emery}},
  \bibinfo{author}{\bibfnamefont{Y.}~\bibnamefont{Hishikawa}},
  \bibnamefont{and} \bibinfo{author}{\bibfnamefont{W.}~\bibnamefont{Warta}},
  \bibinfo{journal}{Prog. Photovolt: Res. Appl.} \textbf{\bibinfo{volume}{19}},
  \bibinfo{pages}{84} (\bibinfo{year}{2011}).

\bibitem[{\citenamefont{Green et~al.}(2012)\citenamefont{Green, Emery,
  Hishikawa, Warta, and Dunlop}}]{Green2012}
\bibinfo{author}{\bibfnamefont{M.~A.} \bibnamefont{Green}},
  \bibinfo{author}{\bibfnamefont{K.}~\bibnamefont{Emery}},
  \bibinfo{author}{\bibfnamefont{Y.}~\bibnamefont{Hishikawa}},
  \bibinfo{author}{\bibfnamefont{W.}~\bibnamefont{Warta}}, \bibnamefont{and}
  \bibinfo{author}{\bibfnamefont{E.~D.} \bibnamefont{Dunlop}},
  \bibinfo{journal}{Prog. Photovolt: Res. Appl.} \textbf{\bibinfo{volume}{20}},
  \bibinfo{pages}{12} (\bibinfo{year}{2012}).

\bibitem[{\citenamefont{Deibel and Dyakonov}(2010)}]{Deibel2010}
\bibinfo{author}{\bibfnamefont{C.}~\bibnamefont{Deibel}} \bibnamefont{and}
  \bibinfo{author}{\bibfnamefont{V.}~\bibnamefont{Dyakonov}},
  \bibinfo{journal}{Rep. Prog. Phys.} \textbf{\bibinfo{volume}{73}},
  \bibinfo{pages}{096401} (\bibinfo{year}{2010}).

\bibitem[{\citenamefont{Brabec et~al.}(2010)\citenamefont{Brabec, Gowrisanker,
  Halls, Laird, Jia, and Williams}}]{Brabec2010}
\bibinfo{author}{\bibfnamefont{C.~J.} \bibnamefont{Brabec}},
  \bibinfo{author}{\bibfnamefont{S.}~\bibnamefont{Gowrisanker}},
  \bibinfo{author}{\bibfnamefont{J.~J.~M.} \bibnamefont{Halls}},
  \bibinfo{author}{\bibfnamefont{D.}~\bibnamefont{Laird}},
  \bibinfo{author}{\bibfnamefont{S.}~\bibnamefont{Jia}}, \bibnamefont{and}
  \bibinfo{author}{\bibfnamefont{S.~P.} \bibnamefont{Williams}},
  \bibinfo{journal}{Adv. Mater.} \textbf{\bibinfo{volume}{22}},
  \bibinfo{pages}{3839} (\bibinfo{year}{2010}).

\bibitem[{\citenamefont{Howard and Laquai}(2010)}]{Howard2010}
\bibinfo{author}{\bibfnamefont{I.~A.} \bibnamefont{Howard}} \bibnamefont{and}
  \bibinfo{author}{\bibfnamefont{F.}~\bibnamefont{Laquai}},
  \bibinfo{journal}{Macromol. Chem. Phys.} \textbf{\bibinfo{volume}{211}},
  \bibinfo{pages}{2063} (\bibinfo{year}{2010}).

\bibitem[{\citenamefont{Deibel et~al.}(2010)\citenamefont{Deibel, Strobel, and
  Dyakonov}}]{Deibel2010c}
\bibinfo{author}{\bibfnamefont{C.}~\bibnamefont{Deibel}},
  \bibinfo{author}{\bibfnamefont{T.}~\bibnamefont{Strobel}}, \bibnamefont{and}
  \bibinfo{author}{\bibfnamefont{V.}~\bibnamefont{Dyakonov}},
  \bibinfo{journal}{Adv. Mater.} \textbf{\bibinfo{volume}{22}},
  \bibinfo{pages}{4097} (\bibinfo{year}{2010}).

\bibitem[{\citenamefont{Bakulin et~al.}(2012)\citenamefont{Bakulin, Rao,
  Pavelyev, van Loosdrecht, Pshenichnikov, Niedzialek, Cornil, Beljonne, and
  Friend}}]{Bakulin2012}
\bibinfo{author}{\bibfnamefont{A.~A.} \bibnamefont{Bakulin}},
  \bibinfo{author}{\bibfnamefont{A.}~\bibnamefont{Rao}},
  \bibinfo{author}{\bibfnamefont{V.~G.} \bibnamefont{Pavelyev}},
  \bibinfo{author}{\bibfnamefont{P.~H.~M.} \bibnamefont{van Loosdrecht}},
  \bibinfo{author}{\bibfnamefont{M.~S.} \bibnamefont{Pshenichnikov}},
  \bibinfo{author}{\bibfnamefont{D.}~\bibnamefont{Niedzialek}},
  \bibinfo{author}{\bibfnamefont{J.}~\bibnamefont{Cornil}},
  \bibinfo{author}{\bibfnamefont{D.}~\bibnamefont{Beljonne}}, \bibnamefont{and}
  \bibinfo{author}{\bibfnamefont{R.~H.} \bibnamefont{Friend}},
  \bibinfo{journal}{Science} \textbf{\bibinfo{volume}{335}},
  \bibinfo{pages}{1340} (\bibinfo{year}{2012}).

\bibitem[{\citenamefont{Nuzzo et~al.}(2010)\citenamefont{Nuzzo, Aguirre,
  Shahid, Gevaerts, Meskers, and Janssen}}]{Nuzzo2010}
\bibinfo{author}{\bibfnamefont{D.~D.} \bibnamefont{Nuzzo}},
  \bibinfo{author}{\bibfnamefont{A.}~\bibnamefont{Aguirre}},
  \bibinfo{author}{\bibfnamefont{M.}~\bibnamefont{Shahid}},
  \bibinfo{author}{\bibfnamefont{V.~S.} \bibnamefont{Gevaerts}},
  \bibinfo{author}{\bibfnamefont{S.~C.~J.} \bibnamefont{Meskers}},
  \bibnamefont{and} \bibinfo{author}{\bibfnamefont{R.~A.~J.}
  \bibnamefont{Janssen}}, \bibinfo{journal}{Adv. Mater.}
  \textbf{\bibinfo{volume}{22}}, \bibinfo{pages}{4321} (\bibinfo{year}{2010}).

\bibitem[{\citenamefont{Liedtke et~al.}(2011)\citenamefont{Liedtke, Sperlich,
  Kraus, Baumann, Deibel, Wirix, Loos, Cardona, and Dyakonov}}]{Liedtke2011}
\bibinfo{author}{\bibfnamefont{M.}~\bibnamefont{Liedtke}},
  \bibinfo{author}{\bibfnamefont{A.}~\bibnamefont{Sperlich}},
  \bibinfo{author}{\bibfnamefont{H.}~\bibnamefont{Kraus}},
  \bibinfo{author}{\bibfnamefont{A.}~\bibnamefont{Baumann}},
  \bibinfo{author}{\bibfnamefont{C.}~\bibnamefont{Deibel}},
  \bibinfo{author}{\bibfnamefont{M.~J.~M.} \bibnamefont{Wirix}},
  \bibinfo{author}{\bibfnamefont{J.}~\bibnamefont{Loos}},
  \bibinfo{author}{\bibfnamefont{C.~M.} \bibnamefont{Cardona}},
  \bibnamefont{and} \bibinfo{author}{\bibfnamefont{V.}~\bibnamefont{Dyakonov}},
  \bibinfo{journal}{J. Am. Chem. Soc.} \textbf{\bibinfo{volume}{133}},
  \bibinfo{pages}{9088} (\bibinfo{year}{2011}).

\bibitem[{\citenamefont{Mort et~al.}(1980)\citenamefont{Mort, Chen, Troup,
  Morgan, Knights, and Lujan}}]{Mort1980}
\bibinfo{author}{\bibfnamefont{J.}~\bibnamefont{Mort}},
  \bibinfo{author}{\bibfnamefont{I.}~\bibnamefont{Chen}},
  \bibinfo{author}{\bibfnamefont{A.}~\bibnamefont{Troup}},
  \bibinfo{author}{\bibfnamefont{M.}~\bibnamefont{Morgan}},
  \bibinfo{author}{\bibfnamefont{J.}~\bibnamefont{Knights}}, \bibnamefont{and}
  \bibinfo{author}{\bibfnamefont{R.}~\bibnamefont{Lujan}},
  \bibinfo{journal}{Phys. Rev. Lett.} \textbf{\bibinfo{volume}{45}},
  \bibinfo{pages}{1348} (\bibinfo{year}{1980}).

\bibitem[{\citenamefont{Offermans et~al.}(2006)\citenamefont{Offermans,
  Meskers, and Janssen}}]{Offermans2006}
\bibinfo{author}{\bibfnamefont{T.}~\bibnamefont{Offermans}},
  \bibinfo{author}{\bibfnamefont{S.~C.~J.} \bibnamefont{Meskers}},
  \bibnamefont{and} \bibinfo{author}{\bibfnamefont{R.~A.~J.}
  \bibnamefont{Janssen}}, \bibinfo{journal}{J. Appl. Phys.}
  \textbf{\bibinfo{volume}{100}}, \bibinfo{pages}{074509}
  (\bibinfo{year}{2006}).

\bibitem[{\citenamefont{Kniepert et~al.}(2011)\citenamefont{Kniepert, Schubert,
  Blakesley, and Neher}}]{Kniepert2011}
\bibinfo{author}{\bibfnamefont{J.}~\bibnamefont{Kniepert}},
  \bibinfo{author}{\bibfnamefont{M.}~\bibnamefont{Schubert}},
  \bibinfo{author}{\bibfnamefont{J.~C.} \bibnamefont{Blakesley}},
  \bibnamefont{and} \bibinfo{author}{\bibfnamefont{D.}~\bibnamefont{Neher}},
  \bibinfo{journal}{Phys. Chem. Lett.} \textbf{\bibinfo{volume}{2}},
  \bibinfo{pages}{700} (\bibinfo{year}{2011}).

\bibitem[{\citenamefont{Kern et~al.}(2011)\citenamefont{Kern, Schwab, Deibel,
  and Dyakonov}}]{Kern2011}
\bibinfo{author}{\bibfnamefont{J.}~\bibnamefont{Kern}},
  \bibinfo{author}{\bibfnamefont{S.}~\bibnamefont{Schwab}},
  \bibinfo{author}{\bibfnamefont{C.}~\bibnamefont{Deibel}}, \bibnamefont{and}
  \bibinfo{author}{\bibfnamefont{V.}~\bibnamefont{Dyakonov}},
  \bibinfo{journal}{Phys. Status Solidi RRL} \textbf{\bibinfo{volume}{5}},
  \bibinfo{pages}{364} (\bibinfo{year}{2011}).

\bibitem[{\citenamefont{Hallermann et~al.}(2008)\citenamefont{Hallermann,
  Haneder, and {Da Como}}}]{Hallermann2008}
\bibinfo{author}{\bibfnamefont{M.}~\bibnamefont{Hallermann}},
  \bibinfo{author}{\bibfnamefont{S.}~\bibnamefont{Haneder}}, \bibnamefont{and}
  \bibinfo{author}{\bibfnamefont{E.}~\bibnamefont{{Da Como}}},
  \bibinfo{journal}{Appl. Phys. Lett.} \textbf{\bibinfo{volume}{93}},
  \bibinfo{pages}{053307} (\bibinfo{year}{2008}).

\bibitem[{\citenamefont{Shrotriya et~al.}(2006)\citenamefont{Shrotriya, Li,
  Yao, Moriarty, Emery, and Yang}}]{Shrotriya2006}
\bibinfo{author}{\bibfnamefont{V.}~\bibnamefont{Shrotriya}},
  \bibinfo{author}{\bibfnamefont{G.}~\bibnamefont{Li}},
  \bibinfo{author}{\bibfnamefont{Y.}~\bibnamefont{Yao}},
  \bibinfo{author}{\bibfnamefont{T.}~\bibnamefont{Moriarty}},
  \bibinfo{author}{\bibfnamefont{K.}~\bibnamefont{Emery}}, \bibnamefont{and}
  \bibinfo{author}{\bibfnamefont{Y.}~\bibnamefont{Yang}},
  \bibinfo{journal}{Adv. Funct. Mater.} \textbf{\bibinfo{volume}{16}},
  \bibinfo{pages}{2016} (\bibinfo{year}{2006}).

\bibitem[{\citenamefont{Deibel}(2009)}]{Deibel2009a}
\bibinfo{author}{\bibfnamefont{C.}~\bibnamefont{Deibel}},
  \bibinfo{journal}{Phys. Status solidi A} \textbf{\bibinfo{volume}{2736}}
  (\bibinfo{year}{2009}), ISSN \bibinfo{issn}{18626300}.

\bibitem[{\citenamefont{Limpinsel et~al.}(2010)\citenamefont{Limpinsel,
  Wagenpfahl, Mingebach, and Deibel}}]{Limpinsel2010}
\bibinfo{author}{\bibfnamefont{M.}~\bibnamefont{Limpinsel}},
  \bibinfo{author}{\bibfnamefont{A.}~\bibnamefont{Wagenpfahl}},
  \bibinfo{author}{\bibfnamefont{M.}~\bibnamefont{Mingebach}},
  \bibnamefont{and} \bibinfo{author}{\bibfnamefont{C.}~\bibnamefont{Deibel}},
  \bibinfo{journal}{Phys. Rev. B} \textbf{\bibinfo{volume}{81}},
  \bibinfo{pages}{085203} (\bibinfo{year}{2010}).

\bibitem[{\citenamefont{Pivrikas et~al.}(2005)\citenamefont{Pivrikas,
  Ju\v{s}ka, Mozer, Scharber, Arlauskas, Sariciftci, Stubb, and
  \"{O}sterbacka}}]{Pivrikas2005}
\bibinfo{author}{\bibfnamefont{A.}~\bibnamefont{Pivrikas}},
  \bibinfo{author}{\bibfnamefont{G.}~\bibnamefont{Ju\v{s}ka}},
  \bibinfo{author}{\bibfnamefont{A.}~\bibnamefont{Mozer}},
  \bibinfo{author}{\bibfnamefont{M.}~\bibnamefont{Scharber}},
  \bibinfo{author}{\bibfnamefont{K.}~\bibnamefont{Arlauskas}},
  \bibinfo{author}{\bibfnamefont{N.}~\bibnamefont{Sariciftci}},
  \bibinfo{author}{\bibfnamefont{H.}~\bibnamefont{Stubb}}, \bibnamefont{and}
  \bibinfo{author}{\bibfnamefont{R.}~\bibnamefont{\"{O}sterbacka}},
  \bibinfo{journal}{Phys. Rev. Lett.} \textbf{\bibinfo{volume}{94}},
  \bibinfo{pages}{1} (\bibinfo{year}{2005}).

\bibitem[{\citenamefont{Deibel et~al.}(2009)\citenamefont{Deibel, Wagenpfahl,
  and Dyakonov}}]{Deibel2009}
\bibinfo{author}{\bibfnamefont{C.}~\bibnamefont{Deibel}},
  \bibinfo{author}{\bibfnamefont{A.}~\bibnamefont{Wagenpfahl}},
  \bibnamefont{and} \bibinfo{author}{\bibfnamefont{V.}~\bibnamefont{Dyakonov}},
  \bibinfo{journal}{Phys. Rev. B} \textbf{\bibinfo{volume}{80}},
  \bibinfo{pages}{1} (\bibinfo{year}{2009}).

\bibitem[{\citenamefont{Dennler et~al.}(2006)\citenamefont{Dennler, Mozer,
  Ju\v{s}ka, Pivrikas, \"{O}sterbacka, Fuchsbauer, and
  Sariciftci}}]{Dennler2006}
\bibinfo{author}{\bibfnamefont{G.}~\bibnamefont{Dennler}},
  \bibinfo{author}{\bibfnamefont{A.~J.} \bibnamefont{Mozer}},
  \bibinfo{author}{\bibfnamefont{G.}~\bibnamefont{Ju\v{s}ka}},
  \bibinfo{author}{\bibfnamefont{A.}~\bibnamefont{Pivrikas}},
  \bibinfo{author}{\bibfnamefont{R.}~\bibnamefont{\"{O}sterbacka}},
  \bibinfo{author}{\bibfnamefont{A.}~\bibnamefont{Fuchsbauer}},
  \bibnamefont{and} \bibinfo{author}{\bibfnamefont{N.~S.}
  \bibnamefont{Sariciftci}}, \bibinfo{journal}{Org. Electron.}
  \textbf{\bibinfo{volume}{7}}, \bibinfo{pages}{229} (\bibinfo{year}{2006}).

\bibitem[{\citenamefont{Mozer et~al.}(2005)\citenamefont{Mozer, Dennler,
  Sariciftci, Westerling, Pivrikas, \"{O}sterbacka, and Ju\v{s}ka}}]{Mozer2005}
\bibinfo{author}{\bibfnamefont{A.}~\bibnamefont{Mozer}},
  \bibinfo{author}{\bibfnamefont{G.}~\bibnamefont{Dennler}},
  \bibinfo{author}{\bibfnamefont{N.}~\bibnamefont{Sariciftci}},
  \bibinfo{author}{\bibfnamefont{M.}~\bibnamefont{Westerling}},
  \bibinfo{author}{\bibfnamefont{A.}~\bibnamefont{Pivrikas}},
  \bibinfo{author}{\bibfnamefont{R.}~\bibnamefont{\"{O}sterbacka}},
  \bibnamefont{and}
  \bibinfo{author}{\bibfnamefont{G.}~\bibnamefont{Ju\v{s}ka}},
  \bibinfo{journal}{Phys. Rev. B} \textbf{\bibinfo{volume}{72}},
  \bibinfo{pages}{1} (\bibinfo{year}{2005}).

\bibitem[{\citenamefont{Koster et~al.}(2006)\citenamefont{Koster, Mihailetchi,
  and Blom}}]{Koster2006b}
\bibinfo{author}{\bibfnamefont{L.~J.~A.} \bibnamefont{Koster}},
  \bibinfo{author}{\bibfnamefont{V.~D.} \bibnamefont{Mihailetchi}},
  \bibnamefont{and} \bibinfo{author}{\bibfnamefont{P.~W.~M.}
  \bibnamefont{Blom}}, \bibinfo{journal}{Appl. Phys. Lett.}
  \textbf{\bibinfo{volume}{88}}, \bibinfo{pages}{052104}
  (\bibinfo{year}{2006}).

\bibitem[{\citenamefont{Baumann et~al.}(2008)\citenamefont{Baumann, Lorrmann,
  Deibel, and Dyakonov}}]{Baumann2008}
\bibinfo{author}{\bibfnamefont{A.}~\bibnamefont{Baumann}},
  \bibinfo{author}{\bibfnamefont{J.}~\bibnamefont{Lorrmann}},
  \bibinfo{author}{\bibfnamefont{C.}~\bibnamefont{Deibel}}, \bibnamefont{and}
  \bibinfo{author}{\bibfnamefont{V.}~\bibnamefont{Dyakonov}},
  \bibinfo{journal}{Appl. Phys. Lett.} \textbf{\bibinfo{volume}{93}},
  \bibinfo{pages}{252104} (\bibinfo{year}{2008}).

\end{thebibliography}
\end{document}